# BREAKING MIGNOTTE'S SEQUENCE BASED SECRET SHARING SCHEME USING SMT SOLVER

K. Vishnu Priyanka[1], M. Gowthami[1], O. Susmitha[1], G. Prathyusha[1], Naresh Babu Muppalaneni[2,*]

[1]Student, Department of CSSE, Sree Vidyanikethan Engineering College (Autonomous), India
[2]Associate Professor, Department of CSSE, Sree Vidyanikethan Engineering College (Autonomous), India.

*ABSTRACT*

*The secret sharing schemes are the important tools in cryptography that are used as building blocks in many secured protocols. It is a method used for distributing a secret among the participants in a manner that only the threshold number of participants together can recover the secret and the remaining set of participants cannot get any information about the secret. Secret sharing schemes are absolute for storing highly sensitive and important information. In a secret sharing scheme, a secret is divided into several shares. These shares are then distributed to the participants' one each and thus only the threshold (t) number of participants can recover the secret. In this paper we have used Mignotte's Sequence based Secret Sharing for distribution of shares to the participants. A (k, m) Mignotte's sequence is a sequence of pair wise co-prime positive integers. We have proposed a new method for reconstruction of secret even with t-1 shares using the SMT solver.*

## 1. INTRODUCTION

Secret Sharing Scheme (SSS) is a method in which a key can be divided into n pieces of information called shares such that i) key can be reconstructed from certain authorized groups of shares and ii) key cannot be reconstructed from unauthorized groups of shares[1-3]. There were security threats to data before the discovery of secret sharing schemes, as we needed to keep the various duplicates of information at better places. With the goal that we could recover the information from another source if the information is lost or harmed physically or decimated by an infection or the information is changed at one of the spots. In any case, we realize that the security threats are increased by increasing the copies of data. Now, secret sharing schemes came into existence to address these issues. A secret sharing scheme is a method used by the dealer to distribute the shares to parties in a way that only the authorized set of parties can reconstruct the secret. The secret sharing schemes are the important tools in cryptography that are used as building blocks in many secured protocols, e.g., threshold cryptography, access control, attribute-based encryption. In a secret sharing scheme a dealer will have a secret, there will be n parties, among them any 't' participants are considered as authorized set who can reconstruct the secret. 't' is known as threshold value. The secret cannot be reconstructed even with 't-1' shares[4].

Blakley[5] proposed a SSS which utilizes hyper plane geometry to solve the secret sharing problem. To execute this (t, n) threshold scheme, each one among the n participants is given a hyper-plane equation in a t dimensional space over a finite field. Then, each hyper plane passes through a specific point. The intersection point purpose of the hyper planes is the secret.
If the set of possible shares are similar to that of the secrets and each participant is given just one share, then the secret sharing scheme is said to be ideal.

ɣ= loglYl / loglXl = 1

where X is possible secrets set, Y is possible shares set and is the information rate.

## 1.1 SMT Solver

SMT solver is Satisfiability modulo theory where a decision problem for logical formulas with respect to combinations of background theories expressed in classical first order logic with equality. It can be considered as a form of the constraint satisfaction problem. Generally, SMT solver are useful both for verification, proving for correctness of the program and software testing. Some of the SMT solvers are SIMPLIFY, CVC3, PARADOX, Z3. Here, now we are using Z3 solver for reconstructing the secret.

Z3 is another SMT solver from Microsoft Research. It is focused at solving problems that arise in software verification and software analysis. Consequently, it integrates support for a variety of theories. Z3 uses novel calculations for quantifier instantiation [6] and hypothesis combination [7]. Z3 is a state-of-the art theorem prover from Microsoft Research. It can be utilized to check the satisfiability of logical formulas over one or more theories. Z3 offers a convincing match for software analysis and verification tools, since a few regular software constructs map directly into supported theories.

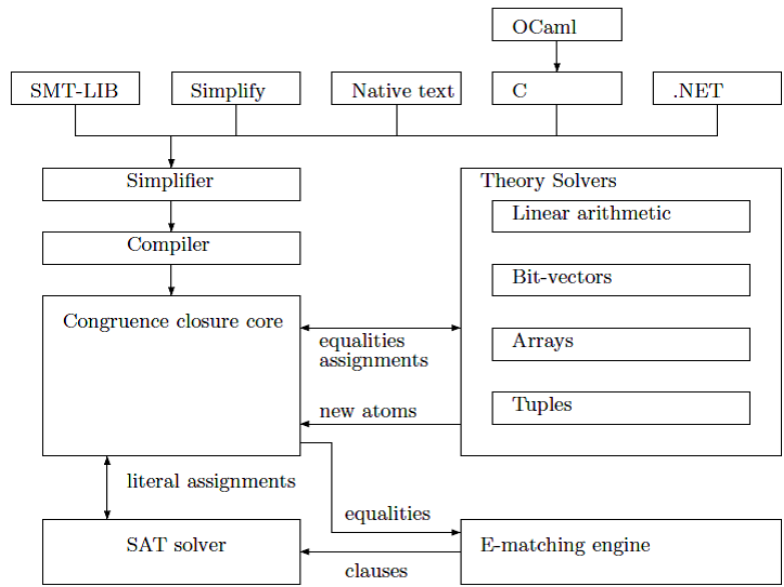

Figure 1: Z3 system architecture

## 1.2 Mignotte's sequence based secret sharing scheme

Assume t and n which represents the threshold value and the number of participants respectively, be two values such that m > 2 and 2 ≤ k ≤ m. A (k, m) Mignotte's sequence is a sequence of pair wise co-prime positive integers $M_1 < M_2 < ... < M_m$ such that $\prod_{i=1}^{k} M_i$, the product of the smallest k of them is greater than the product of the

(k-1) biggest one.
For example a (3, 5) Mignotte's sequence is 7,8,9,10,11.

## 2. METHODOLOGY

Secret sharing scheme is a method in which a secret is distributed among 'n' members such that a threshold 't' number of participants can retrieve the secret. We are a proposing a new method in which a secret is revealed for even 't-1' participants using Z3 SMT solver. We are using Z3py API for reconstructing the secret.

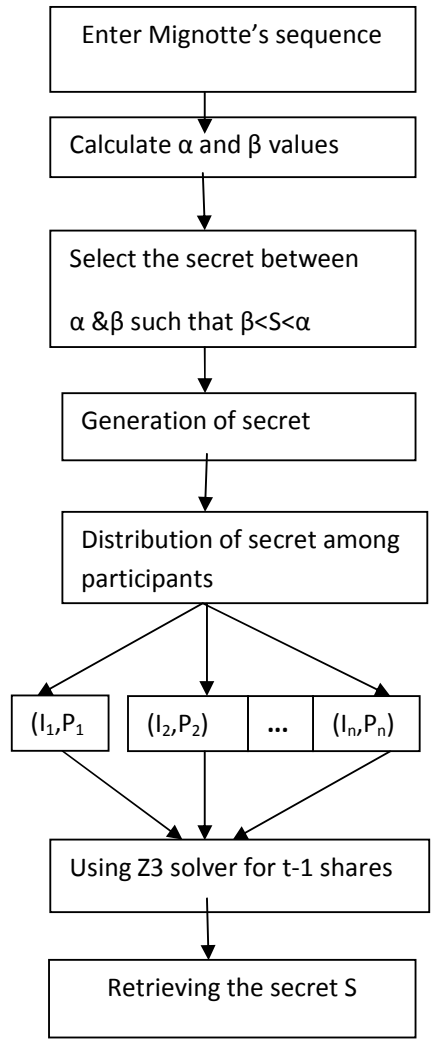

Figure 2: Workflow diagram

### 2.1 Generation

Select some random co-prime 'n' participants. (Eg. 7, 8, 9, 10, 11) Here n=5.Select a threshold value 't' to reconstruct the secret such that t<n.(Eg.3<5).Calculate α and β values where $\alpha = P_1 * P_2 * \ldots * P_t$ and $\beta = P_n * \ldots P_{n-t+3} * P_{n-t+2}$. Select a Secret 'S' between alpha and beta such that $\beta < S < \alpha$. Code is given in Annexure-1.

## 2.2 Distribution

Shares are calculated and assigned among the participants $S_i = K \mod P_i$, $1 \leq i \leq n$.
Where $S_i$ is share for $i^{th}$ participant, K is the secret need to distribute, $P_i$ belongs to mignotte's sequence.

For example let Number of participants (n) be 5. (7, 9, 11, 13, 15). Let t=3 now calculate $\alpha = 7*9*11 = 693$ and $\beta = 13*15 = 195$

Secret should be in between beta and alpha.
Let S be 450 hence it satisfies the condition $\beta < K < \alpha$.
Now calculate the shares $S_i = K \mod P_i$. $(1 \leq i \leq n)$
$S_1 = 450 \mod 7 = 2$
$S_2 = 450 \mod 9 = 0$
$S_3 = 450 \mod 11 = 10$
$S_4 = 450 \mod 13 = 8$
$S_5 = 450 \mod 15 = 0$

Assign the shares $S_i$ to $i^{th}$ participant.
Now the shares are (2,7),(0,9)(10,11)(8,13)(0,15)

## 2.3 Reconstruction

According to Shamir's secret sharing scheme for reconstruction of the Secret we need of 't' shares where 't' is the threshold. But here we are presenting a new approach for reconstruction of secret with 't-1' shares using SMT solver. The SMT solver takes input in smt2lib format. We have generated the constraints as logical formulas using the python code to give as input to SMT solver. The SMT solver checks for the satisfiability of logical formulas over one or more theories.

## 3. RESULTS

C:\z3-4.3.2-x86-win\z3-4.3.2-x86-win\bin>python secret_sharing_scheme.py

Enter Number of Participants:5
Enter Threshold:3
Enter p1:7
Enter p2:9
Enter p3:11
Enter p4:13
Enter p5:17

Enter Secret (between 221 and 693):**330**
Select 't' shares to reconstruct the secret where t=3.
Enter the Participants id for reconstruction:

Enter Partcipant id(0-4):1
Enter Partcipant id(0-4):2
Enter Partcipant id(0-4):4

C:\z3-4.3.2-x86-win\z3-4.3.2-x86-win\bin>python sss.py

**[S = 330, I2 = 7, I1 = 0, I0 = 6]**
[S = 2013, I1 = 0, I0 = 6, I2 = 7]
[S = 3696, I1 = 0, I0 = 6, I2 = 7]
[S = 5379, I1 = 0, I0 = 6, I2 = 7]
[S = 7062, I1 = 0, I0 = 6, I2 = 7]

Select (t-1) shares for reconstruction
C:\z3-4.3.2-x86-win\z3-4.3.2-x86-win\bin>python sss.py
[S = 33, I1 = 0, I0 = 6]
[S = 132, I1 = 0, I0 = 6]
[S = 231, I1 = 0, I0 = 6]
**[S = 330, I1 = 0, I0 = 6]**
[S = 429, I1 = 0, I0 = 6]

The SMT solver is verifying for the satisfiability. If solution exists by satisfying the constraints, the solution will be displayed. The SMT solver is trying to give all the possible solutions for the given constraints. Finally we got the secret S= 330 with 2 shares.

## 4. CONCLUSION

Now-a-days keeping any information secret and safe has become a very important thing for everyone. Secret sharing schemes are the methods that are necessary for this purpose to keep the secret confidential. In a secret sharing scheme, a secret is divided into several shares. These shares are then distributed to the participants' one each and thus only the threshold (t) number of participants can recover the secret. In this paper we made an attempt for reconstruction of secret with 't-1' shares using SMT solver. The constraints in the form of logical formulas are given as the input to SMT solver to check for the satifiability. Hence new secret sharing schemes have to evolve to protect against these kind of attacks.

**Annexure-1**
**# Reading of Mignotte Sequence**
```
def mig_sequence():
        for i in range(n):
                if(i==0):
                        P[i]=input("Enter p{}:".format(i+1))
                if(i>0):
                        # pi,pi-1 are not co primes read pi again
                        while(1):
                                P[i]=input("Enter p{}:".format(i+1))
                                if(int(P[i])<=int(P[i-1])):
                                        continue;
                                if(coprime(int(P[i]),int(P[i-1]))):
                                        break;
```

**#reading mig sequence and calculate alpha and beta and verify  beta is lessthan alpha**

```
while(1):

mig_sequence()
        alpha=1
        for i in range(t):
                alpha=alpha*P[i]
        print alpha

        beta=1
        for i in range(t-1):
                beta=beta*P[n-i-1]
        print beta
        if(beta<alpha):
                break;
        else:
                print("The given input is not mignotte seqeunce\nPlease re enter the sequence\n")
```

**#reading the secret such that beta<Secret<alpha**

```
while(1):
        S=input("Enter Secret (between {} and {}):".format(beta,alpha))
        if(S<beta or S>alpha):
                print("Invalid Secret")
        else:
                break;
```

**# Shares Distribution**

```
target = open("secret.txt", 'w')
target.truncate()
for i in range(n):
        target.write("{} , {} \n".format(S%P[i],P[i]))
target.close();
```

```
#Secret Reconstruction
#reading line by line and storing in array
with open("secret.txt", "r") as ins:
    array = []
    for line in ins:
        array.append(line)
#creating Z3py file
target = open("sss.py", 'w')
target.write("from z3 import *\n")
I = ["I{}".format(i) for i in range(t)]
for i in range(t):
        target.write("{}=Int('{}')\n".format(I[i],I[i]))

target.write("S=Int('S')\n")

target.write("s=Solver()\n")
target.write("s.add(S>0)\n")

#input from user for selecting shares
print("Enter the Participants id for reconstruction:\n")
i=0
while i<t:
        p=input("Enter Partcipant id(0-{}):".format(n-1))
        if(p<0 or p>=n):
                i=i-1
                continue

        x=array[p].split( )

        target.write("s.add({}==S%{})\n".format(I[i],x[2]))

        target.write("s.add({}=={})\n".format(I[i],x[0]))
        i=i+1
target.write("while s.check() == sat:\n")
target.write("   print(s.model())\n\tm = s.model()\n\tblock = []\n\tfor d in m:\n\t\tc = d()\n\t\tblock.append(c != m[d])\n\ts.add(Or(block))\n")

target.close()
```

#sss.py
**A sss file is created to take Participants id for reconstruction by using the SMT solver.**

```
from z3 import *
I0=Int('I0')
I1=Int('I1')
S=Int('S')
s=Solver()
s.add(S>0)
s.add(I0==S%9)
s.add(I0==6)
s.add(I1==S%11)
```

```
s.add(I1==0)
while s.check() == sat:
        print(s.model())
        m = s.model()
        block = []
        for d in m:
                c = d()
                block.append(c != m[d])
        s.add(Or(block))
```